\documentclass[11pt]{iopart}
\usepackage{iopams}
\usepackage{graphicx}
\eqnobysec

\newcommand{\be}{\begin{equation}}
\newcommand{\ee}{\end{equation}}

\newcommand{\ba}{\begin{array}}
\newcommand{\ea}{\end{array}}
\newcommand{\bea}{\begin{eqnarray}}
\newcommand{\eea}{\end{eqnarray}}

\newcommand{\nn}{\nonumber\\ }

\begin{document}
\title{Infinite families of superintegrable systems separable in subgroup coordinates}
\author{ Daniel  L\'evesque, Sarah Post and Pavel Winternitz}

\begin{abstract}
A method is presented that makes it possible to embed a subgroup separable superintegrable system into an infinite family of systems that are integrable and exactly-solvable. It is shown that in two dimensional Euclidean  or pseudo-Euclidean spaces the method also preserves superintegrability. Two infinite families of classical and quantum superintegrable systems are obtained in two-dimensional pseudo-Euclidean space whose classical trajectories and quantum eigenfunctions are investigated. In particular, the wave-functions are expressed in terms of Laguerre and generalized Bessel polynomials. 
\end{abstract}

\section{Introduction}
The purpose of this paper is to show how infinite families of superintegrable systems are obtained from individual known ones which allow separation of variables in at least one subgroup type coordinate system. In this paper we restrict to two-dimensional spaces that are real forms of the complex Euclidean space $E_2(\mathbb{C}).$ The systems obtained in real Euclidean space are already well known \cite{ EvansVerrier2008, JauchHill1940,PW20101,   RTW2008,  TTW2009}. We also obtain two infinite families of superintegrable and exactly-solvable systems on the pseudo-Euclidean plane, $E_{1,1}$, previously obtained in \cite{kalnins2010tools}. These systems are related by coupling constant metamorphosis and we investigate their classical and quantum systems in depth. 

We recall that a superintegrable system is one that allows more integrals of motion than degrees of freedom. The number of functionally independent classical integrals in an $n$-dimensional space is at most $2n-1$. A superintegrable system in two-dimensions will allow precisely 3 independent integrals. 

Superintegrable systems are rare. As a matter of fact, Bertrand's Theorem \cite{Bertrand, Goldstein} tells us that the only spherically symmetric superintegrable potentials are $V(r)=\alpha/r$ or $V(r)=\alpha r^2$ (in any dimension). On the other hand, these are the two most important potentials in physics. 

Superintegrable systems in classical mechanics are of interest mainly  because, in these systems, all bounded trajectories are closed and the motion is periodic \cite{nekhoroshev1972action}. In quantum mechanics, the energy levels are degenerate and superintegrable systems have been conjectured to be exactly solvable \cite{TempTW}. We recall that in quantum mechanics a system is called exactly solvable if its entire energy spectrum can be calculated algebraically and its wave function are polynomials in the appropriate variables, multiplied by an overall gauge factor (the ground state wave function) \cite{ gonzalez1994quasi,  TempTW, Turbiner1988}.  Both in classical and quantum mechanics, the integrals of motion for superintegrable systems form interesting non-Abelian algebras (under Poisson or Lie commutations, respectively). The algebras are polynomial ones \cite{Dask2001, granovskii1991quadratic, letourneau1995superintegrable,  marquette2009painleve} that exceptionally simplify to Lie or Kac-Moody algebras \cite{daboul1993hydrogen}. 

Until recently, most studies of superintegrable systems concentrated on quadratic superintegrability, i.e. the case where all integrals of motion are at most second-order in the momenta, see e.g \cite{FMSUW,  KKMW, KKM20041, KKM20052, MSVW1967}. Second-order superintegrability is  related to multi-separability in the Hamilton-Jacobi and Schr\"odinger equations. Indeed, in real Euclidean spaces, superintegrability and multi-separability are in one-to-one correspondence. 

It was shown in \cite{FMSUW} that exactly four families of quadratically superintegrable systems exist in $E_2(\mathbb{R})$. One of them allows separation of variables in two subgroup type coordinates, namely
\bea \label{H1} H_1&=&p_r^2+\frac{1}{r^2}p_\theta^2 +\omega^2 r^2+\frac{\alpha}{r^2\cos^2 \theta}+\frac{\beta}{r^2\sin^2\theta}\\
&=& p_x^2+p_y^2+\omega^2\left(x^2+y^2\right)+\frac{\alpha}{x^2}+\frac{\beta}{y^2}\nonumber\eea
(polar and Cartesian coordinates), two in one subgroup and one non-subgroup type, namely
\bea \label{H2} H_2=p_x^2+p_y^2+\omega^2(x^2+4y^2)+\beta y+\frac{\alpha}{x},\eea
(cartesian and parabolic) and 
\bea \label{H3} H_3=p_r^2+\frac{1}{r^2}p_\theta^2+\frac{\alpha}{r}+\frac{1}{r^2}\left(\frac{\beta}{\cos^2 \theta}+\frac{\gamma}{\sin^2 \theta}\right), \eea
(polar and parabolic). 
The fourth system allows separation in two different parabolic systems and is not relevant to this article. 

\section{Adding a parameter to integrable systems that separate in subgroup-type coordinates}
Let us first consider the two-dimensional Euclidean space $E_2(\mathbb{C})$ with its isometry group $E(2, \mathbb{C})$ and its Lie algebra $e(2, \mathbb{C}) \sim \{ L_3, P_1, P_2\}.$ This algebra has three non-equivalent one-dimensional subalgebras, $\{P_1\}$, $\{P_1+iP_2\}$ and $\{ L_3\}.$  Diagonalizing $P_1$, $P_1+iP_2$ or $L_3$ simultaneously with the Laplace operator $\Delta=P_1^2+P_2^2$ corresponds to the introduction of subgroup coordinates, namely cartesian $\{u,v\}$, light-cone $\{\xi=u+v, \eta=u-v\}$ and polar $\{\rho=\sqrt{u^2+v^2}, \theta=arctan\frac u v \}$ coordinates, respectively. 

A classical Hamiltonian on $E_2(\mathbb{C})$ separable in orthogonal subgroup type coordinates (i.e. Cartesian or polar coordinates) will have the following form 
\be\label{Hrealsub} H=p_1^2+f_1(q_1)+\psi(q_1)X, \qquad X=p_2^2+f_2(q_2),\ee
where $(q_1, q_2), (p_1, p_2)$ are the classical coordinates and the conjugate momenta on four-dimensional phase space. In these coordinates, the kinetic energy, or free Hamiltonian, is 
\be \Delta=p_1^2+\psi(q_1)p_2^2. \ee
and $V=f_1(q_1)+\psi(q_1)f_2(q_2)$ is the potential. In $E_2(\mathbb{C})$, the only two inequivalent choices for $\psi$ are  
\be \psi=\left\{\ba{ll}1 & \mbox{ Cartesian coordinates}\\
  1/q_1^2 & \mbox{ polar  coordinates}\ea \right. \ee
   There is a natural quantized version of \eref{Hrealsub} obtained by taking $(q_1, q_2)$ as the position operator and $(p_1,p_2)$ as the momentum one, though there may be quantum corrections to \eref{Hrealsub}. 

We observe immediately that system \eref{Hrealsub} is integrable and that $X$ is an integral of motion. Further, if we introduce a parameter $k$ by replacing $X\rightarrow k^2 X$, the obtained system remains integrable but the metric associated with the kinetic energy is changed. The system is mapped back into flat space by making an additional change of variables $q_2\rightarrow kq_2$, $p_2\rightarrow (1/k)p_2.$ The new Hamiltonian and integral of motion are 
\be \label{Hkgen} H=p_1^2+f_1(q_1)+\psi(q_1)X, \qquad X=p_2^2+k^2f_2(kq_2). \ee
The kinetic energy is the same as in \eref{Hrealsub}, however, the potential term $f_2$ has acquired a parameter $k$. The Hamiltonian \eref{Hkgen} represents an infinite family of integrable Hamiltonians which are separable in the original coordinate system $(q_1, q_2).$

Let us know apply the above procedure to known two-dimensional superintegrable systems. Starting from the system \eref{H1} in polar coordinates, we arrive at the TTW system
\be\label{HTTW} H_k^{TTW}=p_r^2+\frac{1}{r^2}p_\theta^2+\omega^2r^2 +\frac{k^2}{r^2}\left[\frac{\alpha }{\cos^2 k \theta}+\frac{\beta}{\sin^2 k\theta}\right].\ee
It is integrable and exactly solvable and was conjectured to be superintegrable for rational k  \cite{TTW2009, TTW2010}. Several authors then proved that \eref{HTTW} is superintegrable for all rational k, both in classical and quantum mechanics \cite{ KKM2010quant, KMPTTWClass, QuesneTTWodd} (without any quantum corrections). 

The system \eref{H1} is also separable in Cartesian coordinates and the same procedure applied to both $x$ and $y$ leads to the anisotropic singular (or caged) harmonic oscillator \cite{EvansVerrier2008, RTW2008}
\be\label{Hkanis} H_k^{anis}=p_x^2+p_y^2+\omega^2(k_1^4 x^2+k_2^4y^2)+\frac{\alpha}{x^2}+\frac{\beta}{y^2}.\ee
It is clear that this procedure applied to \eref{H2} in cartesian coordinates will give a subcase of \eref{Hkanis} and in fact \eref{H2} is a member of the infinite family \eref{Hkanis}. 

Finally, starting from the deformed Coulomb system \eref{H3}, we arrive at the system 
\be\label{HPW} H_k^{PW}=p_r^2+\frac{1}{r^2}p_\theta^2+\frac{K}{r}+\frac{\alpha k^2}{r^2\cos^2 k\theta}+\frac{\beta k^2}{r^2\sin^2 k\theta}\ee
which was proven \cite{PW20101} to be superintegrable and equivalent via coupling constant metamorphosis to the TTW system \eref{HTTW} (with $k\rightarrow 2k$). 
Thus, \eref{HTTW}, \eref{Hkanis} and \eref{HPW} are the only infinite families of superintegrable systems, obtained in this manner, in $E_2(\mathbb{C})$ that also exist in the real Euclidean space $E_2=E_2(\mathbb{R}).$

Two further superintegrable systems exist in $E_2(\mathbb{C})$ that are separable in one orthogonal subgroup type system of coordinates (polar coordinates) and one non-subgroup type (hyperbolic coordinates). The corresponding Hamiltonians in Cartesian coordinates are 
\be \label{HMDO} H^{DO}=p_1^2+p_2^2+\omega^2(x_1^2+x_2^2) +\alpha \frac{x_1^2+x_2^2}{(x_1-ix_2)^4}+\beta \frac{1}{(x_1+ix_2)^2}\ee
and 
\be H^{DC}=p_1^2+p_2^2+\frac{K}{\sqrt{x_1^2+x_2^2}} +\alpha\frac{1}{(x_1+ix_2)^2}
+\beta \frac{1}{\sqrt{x_1^2+x_2^2}(x_1-ix_2)}.\ee
These two systems were identified as E7 and E17 in \cite{Eis}. They separate in polar coordinates on $E_2(\mathbb{C})$
\[ x_1+ix_2=re^{i\theta}, \qquad  x_1-ix_2=re^{-i\theta},\]
and we have 
\be H^{DO}=p_r^2+\frac{1}{r^2}p_\theta^2+\omega^2r^2+\frac{1}{r^2}\left(\alpha e^{4i\theta} +\beta e^{2i\theta}\right),\ee
\be H^{DC}=p_r^2+\frac{1}{r^2}p_\theta^2+\frac{K}{r}+\frac{1}{r^2}\left(\alpha e^{2i\theta} +\beta e^{i\theta}\right).\ee
In real Euclidean space, $H^{DO}$ and $H^{DC}$ would be respectively a deformed harmonic oscillator and a deformed Coulomb potential and the deforming potential in both cases is complex. 

Still in complex Euclidean space, let us introduce a parameter putting $\theta \rightarrow k \theta.$ This preserves one second order integral and separation of variables in polar coordinates. Following the procedure outlined above with a scaling of the coupling constants, we obtain the integrable families \cite{KKMtools}
\bea H_k^{DO} =p_r^2+\frac{1}{r^2}p_\theta^2+\omega^2r^2+\frac1{r^2}\left(\alpha e^{4ik\theta}+\beta e^{2ik\theta}\right)\label{HKDO}\\
H_k^{DC}=p_r^2+\frac{1}{r^2}p_\theta^2+\frac{K}{r}+\frac1{r^2}\left(\alpha e^{2ik\theta}+\beta e^{ik\theta}\right)\label{HKDC}.\eea
The two Hamiltonians $H_k^{DO}$ and $H_k^{DC}$ are related by coupling constant metamorphosis \cite{HierGrammStackel, KMPostStackel, Kress2007}. In fact, if we consider the Hamilton-Jacobi for $H_k^{DO}$
\be  p_r^2+\frac{1}{r^2}p_\theta^2+\omega^2r^2+\frac1{r^2}\left(\alpha e^{4ik\theta}+\beta e^{2ik\theta}\right) =E,\ee
and solve the equation for $\omega$, we obtain
\be\frac{1}{r^2}\left(p_r^2+\frac{1}{r^2}p_\theta^2\right)-\frac{E}{r^2}+\frac1{r^4}\left(\alpha e^{4ik\theta}+\beta e^{2ik\theta}\right) =-\omega^2.\label{CCMH} \ee
The change of variables $\rho=r^2/2, \qquad \phi=2\theta$, changes the left-hand side of \eref{CCMH} into the Hamiltonian \eref{HKDC} in the new polar coordinates $\rho, \phi$ with $K=-E/2$ and  the constants $\alpha, \beta$ scaled by $1/4.$

Just as in the previous known examples of infinite families of Hamiltonians, there is no guarantee that the infinite family is superintegrable. The superintegrability of each of the families must be verified. In the following sections, we prove that the system \eref{HKDO} is superintegrable in both the classical and quantum case and  hence so is \eref{HKDC} as a result of the coupling constant metamorphosis.

\section{Infinite families of superintegrable systems in the pseudo-Euclidean plane}
Let us now restrict to the real pseudo-Euclidean plane $E_{1,1}$ and study the system \eref{HKDO}. In cartesian coordinates, we have 
\be \fl H_k^{DO}=p_1^2+p_2^2+\omega^2\left(x_1^2+x_2^2\right)+\frac{1}{x_1^2+x_2^2}\left[\alpha\left(\frac{x_1+ix_2}{x_1-ix_2}\right)^{2k}+\beta\left(\frac{x_1+ix_2}{x_1-ix_2}\right)^{k}\right].\ee
Let us put 
\be x_1=u, \qquad ix_2=v, \qquad u,v\in \mathbb{R}.\ee
We obtain a Hamiltonian in $E_{1,1}$ namely
\be\label{Hk12} \fl H_k=p_u^2-p_v^2+\omega^2(u^2-v^2)+\frac{1}{u^2-v^2}\left(\alpha\left(\frac{u+v}{u-v}\right)^{2k}+\beta\left(\frac{u+v}{u-v}\right)^{k}\right), \qquad \alpha, \beta \in \mathbb{R}.\ee
Without loss of generality, we can assume that $k> 0$ and either $\alpha=\beta=0$ or $\alpha\ne 0;$ indeed $\alpha=0$ is equivalent to $\alpha \ne 0, \beta =0$ with $k\rightarrow k/2.$ In the former case, $u,v$ are allowed to take all real values. In the latter case, we require that $u^2-v^2$ be positive unless $k$ is a positive integer. If $k$ is a positive integer, the requirements are relaxed so that it is only required that $u-v\ne 0.$ 

The Hamiltonian \eref{Hk12} allows separation of variables in pseudo-polar coordinates 
\bea u=r\cosh \gamma, \quad &v=r\sinh \gamma,\nn
r^2=u^2-v^2, \quad & \tanh \gamma=\frac{v}{u},\nonumber\eea
with $r\in \mathbb{R}$ for the "time-like" region of the pseudo-Euclidean plane. In the "space-like" region, we must interchange $u$ and $v$. An equivalent but more convenient set of coordinates in $E_{1,1}$ are modified pseudo-polar coordinates 
\bea \rho=u^2-v^2, \quad \sigma=\frac{u+v}{u-v}, \quad u=\frac{\sigma+1}{2}\sqrt{\frac{\rho}{\sigma}}, \quad v=\frac{\sigma-1}{2}\sqrt{\frac{\rho}{\sigma}},\\
-\infty < \rho < \infty, \quad -\infty < \sigma< \infty, \quad \frac{\rho}{ \sigma}=(u-v)^2\geq 0.\eea
These coordinates cover $E_{1,1}$ with the line $u-v=0$ removed.
The Hamiltonian in these coordinates is 
\be\label{Hk} H_k=4\rho p_\rho^2-4\frac{\sigma^2}{\rho}p_\sigma^2+\omega^2\rho +\frac1{\rho}\left(\alpha \sigma^{2k}+\beta \sigma^{k}\right). \ee
We shall show below that the corresponding Hamiltonian system is superintegrable both in classical and quantum mechanics for all rational values of the constant $k.$ The potential in \eref{Hk} is real and finite for $0< \rho< \infty, $ $0< \sigma< \infty$ and we restrict to this region when either $\alpha$ or $\beta$ is different from zero.

\section{The classical system}
\subsection{Separation of the Hamilton-Jacobi equation. Trajectories} \label{traj}
In this section, we solve for bounded trajectories of the Hamiltonian system \eref{Hk} in the case that $\alpha\ne  0$. The case of $\alpha=\beta=0$ is the harmonic oscillator in $E_{1,1},$ similar to the one in $E_2$ although the energy can take negative values. The  trajectories  can be directly obtained from separation of variables in the coordinates $u,v$ as discussed in \ref{SHO}.

Consider a bounded trajectory in the following regime,
\be 0\leq \rho_1\leq \rho \leq \rho_2, \qquad 0\leq \sigma_1 \leq \sigma \leq \sigma _2.\ee
We write the Hamilton-Jacobi equation for $H_k$ as in \eref{Hk} as
\be 4\rho\left(\frac{\partial}{\partial \rho}S\right)^2-4\frac{\sigma^2}{\rho}\left(\frac{\partial}{\partial \sigma}S\right)^2  +\omega^2\rho+\frac{1}{\rho}\left(\alpha \sigma^{2k}+\beta\sigma^{k}\right)=E,\ee
and separate variables, putting
\be \label{separation}4\rho^2\left(\frac{\partial S_1}{\partial \rho}\right)^2+\omega^2\rho^2-E\rho=4\sigma^2\left(\frac{\partial}{\partial \sigma}S_2\right)^2-\alpha\sigma^{2k}-\beta \sigma^{k}=A\ee
with 
\[ S=S_1(\rho)+S_2(\sigma)-Et, \qquad A\in \mathbb{R}. \]
The functions $S_1$ and $S_2$ are  
\bea \label{S1} S_1(\rho)=
-\frac12\int_{\rho}^{\rho_2}\sqrt{-\omega^2+E\tau^{-1}+A\tau^{-2}}d\tau+c_1, \qquad  \rho>0. \\ \label{S2} 
S_2(\sigma)=
-\frac12\int_{\sigma}^{\sigma_2}\sqrt{A\tau^{-2}+\alpha\tau^{2k-2} + \beta\tau^{k-2}} d\tau+c_2, \qquad \sigma>0.
 \eea
 To calculate the trajectories, we  solve the equations 
\bea \label{Hkdeteqr} \frac{\partial S}{\partial E}=\frac{\partial S_1}{\partial E}-t=\delta_1, \\ \label{Hkdeteqs}\frac{\partial S}{\partial A}=\frac{\partial S_1}{\partial A}+\frac{\partial S_2}{\partial A}=\delta_2\eea
where $\delta_1$ and $\delta_2$ are arbitrary real constants. 
Beginning with \eref{Hkdeteqr}, we obtain
\be\label{detr1} t+\delta_1=
- \frac14\int_{\rho}^{\rho_2} \frac{d\tau}{\sqrt{-\omega^2\tau^2+E\tau+A}}. \ee
Immediately from this equation, we see that unless $\omega^2>0$, the trajectories will be unbounded at $t\rightarrow \infty.$ Further, since the polynomial  $-\omega^2 \rho^2+E\rho+A$ is negative for large $r,$ the integrand \eref{detr1} will take real values if and only if the polynomial has two distinct real roots.
With this requirement, the integrated form of \eref{detr1} is
\bea \label{deteq1}\fl \rho=
\frac{1}{2\omega^2}\left[E-\sqrt{D_1}\sin\left(-4\omega (t+\delta_1)+\arcsin\left(\frac{-2\omega^2 \rho_2+E}{\sqrt{D_1}}\right)\right)\right],\\ D_1=E^2+4\omega^2A>0.\nonumber\eea

From these equations, we see that the limits on the values of $\rho$ are 
\be \label{rbounds} \rho_1=\frac{E-\sqrt{D_1}}{2\omega^2}\leq \rho\leq  \rho_2=\frac{E+\sqrt{D_1}}{2\omega^2}\ee
and hence the equation for $\rho$ simplifies as 
\be\label{rhot} \rho=\frac{1}{2\omega^2}\left[E+\sqrt{D_1}\cos \left(4\omega (t+\delta_1)\right)\right].\ee
Note that that $\rho$ has a period of $ \pi/2\omega$ in $t.$ In the case that $\alpha=\beta=0$, we have the trajectories for the harmonic oscillator, compare with \eref{rhosho}.  In this case  $A$ is positive and so $\rho_1\leq 0\leq \rho_2$. On the other hand, if we assume either $\alpha$ or $\beta$ is non-zero, this leads to the requirement $\rho>0$ which implies both $E\geq 0$ and $A\leq 0$, as well as $E^2+4\omega^2 A\geq 0.$
 
Next, we solve the equation \eref{Hkdeteqs} in the sector where $\rho, \sigma>0$. 
The first relevant integral is, 
\bea \frac{\partial S_1}{\partial A}&=&\frac{-1}4\int_{\rho}^{\rho_2} \frac{d\tau}{\tau^{-2}\sqrt{-\omega^2+E\tau^{-1}+A\tau^{-2}}}
\label{rint}\eea
with
\bea\frac{\partial S_1}{\partial A}&=&
\frac{1}{4\sqrt{-A}}\left[\arcsin\left(\frac{2A\rho^{-1}+E}{\sqrt{D_1}}\right) -\frac{\pi}{2}\right], \qquad A<0,\label{S1A}\eea
The second relevant integral in \eref{Hkdeteqs} is, for $k\ne 0$ 
\bea \frac{\partial S_2}{\partial A}&=&\frac{-1}4\int_{\sigma}^{\sigma_2}\frac{d\tau}{\tau^2\sqrt{A\tau^{-2}+\alpha\tau^{2k-2} + \beta\tau^{k-2}}},\nn
&=& \frac{1}{4k}\int_\sigma^{\sigma_2}\frac{d(\tau^{-k})}{\sqrt{A\tau^{-2k}+\beta \tau^{-k}+\alpha}},\eea
which is integrated as
\bea \fl \frac{\partial S_2}{\partial A}\label{S2A}=
\frac{1}{4k\sqrt{-A}}\left[\arcsin\left(\frac{2A\sigma^{-k}+\beta}{\sqrt{D_2}}\right)- \arcsin\left(\frac{2A\sigma_2^{-k}+\beta}{\sqrt{D_2}}\right)\right],\\
D_2=\beta^2-4\alpha A>0, \qquad \sigma_2=\frac{\beta+\sqrt{D_2}}{2|A|}.\nonumber
\eea
Thus, all bounded trajectories 
satisfy  $E\geq 0$ and $A\leq 0$ with
\be \frac{\beta-\sqrt{D_2}}{2|A|}\leq \sigma_1\leq\sigma^k \leq\sigma_2\leq \frac{\beta+\sqrt{D_2}}{2|A|}, \quad \beta>0, \quad \alpha<0.\ee
The condition $D_2>0$ provides a fundamental restriction on the values of the separation constant, namely 
\be 0< -A\leq\frac{\beta^2}{4\alpha}.\ee

We are now in a position to calculate the trajectories. Let us define new variables 
\be  Z=\frac{2A\rho^{-1}+E}{\sqrt{D_1}}, \qquad W_k=\frac{2A\sigma^{-k}+\beta}{\sqrt{D_2}}\ee
and the constant 
\be \Gamma_k=k\left(4\delta_2\sqrt{-A}+\frac{\pi}{2}\right)+\arcsin\left(\frac{2A(-\beta+\sqrt{D_2})+\beta}{\sqrt{D_2}}\right).\ee
Putting \eref{S1A} and \eref{S2A} into \eref{Hkdeteqs}, we obtain an equation for the trajectories, namely
\be \arcsin W_k+k\arcsin Z=\Gamma_k\label{arcsins} \ee
or equivalently
\be\label{deteq2} \sigma^{-k}=\frac{-1}{2A}\left[\beta -\sqrt{D_2}\sin\left(-k\arcsin Z +\Gamma_k\right)\right].\ee
We already know from \eref{rhot} that $\rho$ and hence $Z$ has period $\tau =\frac{\pi}{2\omega}.$ Since $\arcsin$ is a multivalued function, equation \eref{deteq2} does not prove that $\sigma$ has the same period, or that it is periodic at all. On the other hand, if $k$ is rational, equation \eref{deteq2} can be reduced to a single valued implicit function of $\rho$ and $\sigma$ and this proves that the motion is periodic. Put \be k=p/q, \qquad p,q \in \mathbb{N}.\ee
We rewrite \eref{arcsins} as 
\be \cos C_k=\cos\left[q \arccos W_k+p\arccos Z\right] , \qquad C_k=(p+q)\frac{\pi}{2}-q\Gamma_k.\label{deteqcos}\ee
Making use of the Chebyshev polynomials defined as, 
\be \label{Cheby} T_n(x)=\cos\left(n\arccos(x)\right), \qquad U_n(x)=\frac{\sin\left((n+1)\arccos(x)\right)}{\sin\arccos x}\ee
equation \eref{deteqcos} becomes
\bea\label{cosCk1} cos(C_k)=T_q(W_k)T_p(Z)-U_{q-1}(W_k)U_{p-1}(Z)\sqrt{(W_k^2-1)(Z^2-1)}.\eea
Using the explicit formulas for the Chebyshev polynomials 
 $$ T_{n}(x)=\sum_{j=0}^{\left[n/2\right] } ({}_{2j}^{n}) x^{n-2j}(x^2-1)^j, \qquad  U_{n}(x)=\sum_{j=0}^{\left[ n/2\right]} ({}_{2j+1}^{n+1}) x^{n-2j}(x^2-1)^j $$
where n is a positive integer, we can now write \eref{cosCk1} as
\bea\label{cosCk2}\fl  cos(C_k)=\left(\sum_{j=0}^{\left[q/2\right] } ({}_{2j}^{q}) W_k^{q-2j}(W_k^2-1)^j\right)\left(  \sum_{j=0}^{\left[p/2\right] } ({}_{2j}^{p}) Z^{p-2j}(Z^2-1)^j\right)\\
 \quad \fl -\sqrt{(W_k^2-1)(Z^2-1)}\left(\sum_{j=0}^{\left[ (q-1)/2\right]} ({}_{2j+1}^{q}) W_k^{(q-1)-2j}(W_k^2-1)^{j}\right)\left(\sum_{j=0}^{\left[ (p-1)/2\right]} ({}_{2j+1}^{p}) Z^{(p-1)-2j}(Z^2-1)^{j}\right).\nonumber\eea

Let us summarize the results on the trajectories. Without loss of generality, we have assumed $k>0$ and $\alpha \ne 0.$ 
\begin{enumerate}
\item Bounded trajectories are obtained for  
\be 0>\alpha, 0\leq \beta, \qquad 0< -A \leq \frac{E^2}{4\omega^2}, \qquad 0<-A \leq \frac{\beta^2}{2\alpha}. \label{bounds}\ee
Thus, for this system in $E_{1,1}$, contrary to the case of the real Euclidean plane $E_2$ \cite{TTW2009, TTW2010}, we obtain a fundamental upper bound on the value of the separation constant $A$ \eref{bounds}
\item For rational $k$, all bounded trajectories are periodic.
\end{enumerate}

\subsection{Superintegrability of the classical system for rational k}
The classical trajectories  can be used to explicitly construct an integral of the motion which is polynomial in the momenta (see e.g. \cite{kalnins2010tools, KMPTTWClass, KKM2010JPA, PW20101}).  The additional classical integral of the motion was first constructed in \cite{kalnins2010tools}. It is also possible to construct the integral directly from the bounded trajectories by replacing the conserved quantities $A$ and $E$ by their phase-space counterparts 
\bea E\rightarrow H, \qquad A \rightarrow 4\sigma^2 p_\sigma^2-\alpha \sigma^{2k}-\beta \sigma^{k}.\eea
 The relevant quantities are 
\bea \label{Zp}Z=\frac{4\rho^2p_{\rho}^2+4\sigma^2p_{\sigma}^2+\omega^2\rho^2-\alpha\sigma^{2k}-\beta\sigma^{k}}{\rho\sqrt{D_1}},\\
\label{Wkp} W_k=\frac{8\sigma^{2-k}p_{\sigma}^2-2\alpha\sigma^k-\beta}{\sqrt{D_2}},\\
\label{sqrtp} \sqrt{(1-Z^2)(1-W_k^2)}=\frac{16\sigma^{1-k} p_\rho p_\sigma(\alpha\sigma^{2k}+\beta\sigma^{k}-4\sigma^2p_\sigma^2)}{\sqrt{D_1D_2}}.\eea
\bea D_1=H^2+4\omega^2( 4\sigma^2 p_\sigma^2-\alpha \sigma^{2k}-\beta \sigma^{k})^2\\
D_2=\beta^2-4\alpha( 4\sigma^2 p_\sigma^2-\alpha \sigma^{2k}-\beta \sigma^{k}).\eea
Using the parity properties of the Chebyshev polynomials, the following function on phase space is a constant of the motion which is polynomial in momenta of total degree no greater than $2p+2q$,
\be \fl  \mathcal{L}=\sqrt{D_1}^{p}\sqrt{D_2}^{q}\left[T_q(W_k)T_p(Z)-U_{q-1}(W_k)U_{p-1}(Z)\sqrt{(W_k^2-1)(Z^2-1)}\right].\ee 
Here we see the explicit link between the integrals of motion and the bounded trajectories.
As discussed elsewhere, see e.g. \cite{KMPTTWClass}, this is not necessarily the integral of motion of lowest degree in the momenta. Nevertheless, this integral shows that the Hamiltonian \eref{Hk} is classically superintegrable for rational $k.$

\subsection{Examples of trajectories}

We shall now calculate and plot the trajectories directly from \eref{cosCk1}. For $k=p/q$ integer,\eref{cosCk1} becomes 
\bea cos(C_p)=W_pT_p(Z)-U_{p-1}(Z)\sqrt{(W_p^2-1)(Z^2-1)}\eea
and trajectories for $k$ = 1, 2 and 3 satisfy:
\bea\fl \ba{ll}k=1:& cos(C_1)=W_1Z-\sqrt{(W_1^2-1)(Z^2-1)}\\
 k=2:& cos(C_2)=W_2(2Z^2-1) -2Z\sqrt{(W_2^2-1)(Z^2-1)}\\
 k=3:& cos(C_3)=W_3(4Z^3-3Z) -(4Z^3-1)\sqrt{(W_3^2-1)(Z^2-1)}.\ea\nonumber\eea
 The trajectories are plotted in figure 1 with constants chosen as indicated in accordance with \eref{bounds}

\begin{figure}\label{plot1}
\begin{center}$
\begin{array}{ccc}
\vspace{0pt}
\includegraphics[width=1.9 in]{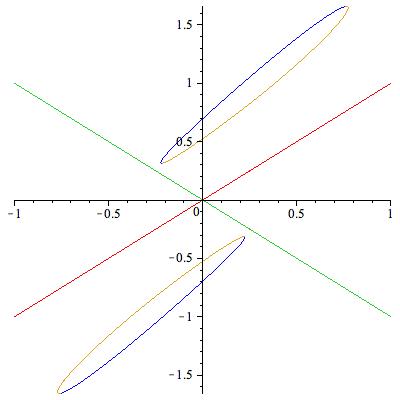} & \includegraphics[width=1.9in]{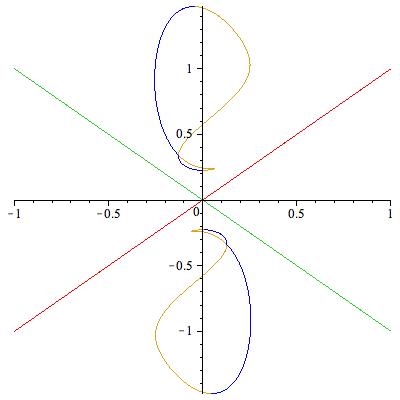} & \includegraphics[width=1.9in]{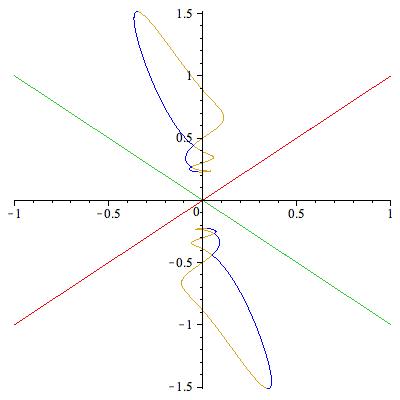} \\
k=1 &  k=2 &  k=3  \end{array}$
\end{center}
\caption{Trajectories of specific cases for $k$ integer. The constants are chosen to be $ A = -1, \alpha = -2, \beta = 6, E = 20, \omega = 3 $ and $ \delta_2 = \pi/32. $ }
\end{figure}

\begin{figure}\label{plot2}
\begin{center}$
\begin{array}{ccc}
\vspace{0pt}
\includegraphics[width=1.9 in]{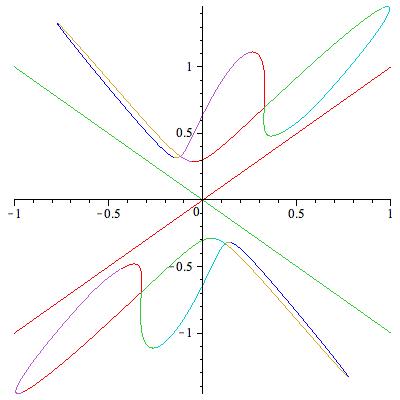} & \includegraphics[width=1.9in]{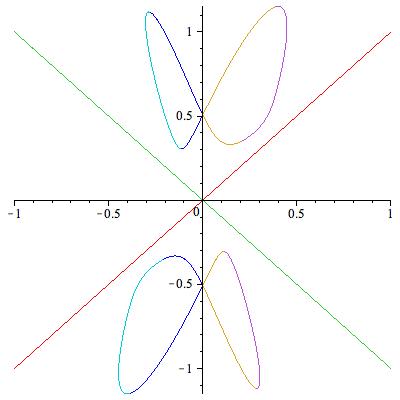} & \includegraphics[width=1.9in]{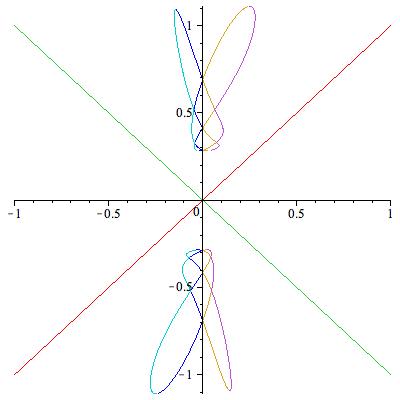} \\
k=1/3, \delta_2 = 3\pi/12  &  k=1/2, \delta_2 = \pi/6  &  k=3/2, \delta_2 = \pi/12   \end{array}$
\end{center}
\caption{Trajectories of specific cases for  rational $k$. The constants are chosen to be $ A = -3/2, \alpha = -1, \beta = 3, E = 20  $ and $\omega = 4. $ }
\end{figure}

The trajectories  for $k$ = 1/3, 1/2 and 3/2 satisfy: 
\bea\fl  \ba{ll} k=\frac 1 3: & \cos(C_{1/3})=(4W_{1/3}^3-3W_{1/3})Z-(4W_{1/3}^2-1)\sqrt{(W_{1/3}^2-1)(Z^2-1)}\\
 k=\frac 1 2 :& \cos(C_{1/2})=(2W_{1/2}^2-1)Z -2W_{1/2}\sqrt{(W_{1/2}^2-1)(Z^2-1)} \\
 k=\frac 3 2: & \cos(C_{3/2})=(2W_{3/2}^2-1)(4Z^3-3Z)-2W_{3/2}(4Z^2-1)\sqrt{(W_{3/2}^2-1)(Z^2-1)}.\ea \nonumber \eea
 The trajectories are plotted in  figure 2 with constants chosen as indicated in accordance with \eref{bounds}

\section{The quantum system}
In quantum mechanics, the Hamiltonian on pseudo-Euclidean space $E(1,1)$ is 
\be\label{HkM}\fl H_k=-4\rho\frac{\partial^2 }{\partial \rho^2}-4\frac{\partial}{\partial \rho}+\frac{4\sigma^2}{\rho}\frac{\partial^2}{\partial \sigma^2}+\frac{4\sigma}{\rho}\frac{\partial}{\partial \sigma}+\omega^2\rho+\frac{1}{\rho}\left(\alpha\sigma^{2k}+\beta\sigma^k\right).\ee
The integral associated with separation of variables is 
\be L_1=-4{\sigma^2}\frac{\partial^2}{\partial \sigma^2}-4{\sigma}\frac{\partial}{\partial \sigma}-\alpha\sigma^{2k}-\beta\sigma^k.\ee
Here again, we assume that $k>0$ and $\alpha$ is not zero.

\subsection{Wave functions and energy spectrum} \label{52}
We look for wavefunctions which are also eigenfunctions of the integral $L_1$ and hence allow separation of variables
\be\label{Schr} H_k\Psi=E\Psi, \qquad L_1\Psi=A\Psi, \qquad \Psi=R(\rho)S(\sigma).\ee
The functions $R(\rho)$ and $S(\sigma)$ satisfy the following equations
\bea\left[ 4\rho\frac{\partial^2}{\partial \rho^2} +4\frac{\partial}{\partial \rho}-\omega^2 \rho+\frac{A}{\rho}+E\right] R(\rho)=0,\label{detR}\\
\left[4\sigma^2\frac{\partial^2}{\partial \sigma^2} +4\sigma\frac{\partial}{\partial \sigma} +\alpha \sigma^{2k}+\beta \sigma^{k}+A\right]S(\sigma)=0,\label{detS}
\eea
respectively. 

Equation \eref{detR} can be solved in terms of the confluent hypergeometric function and indeed we have two  solutions
\be R^{\pm}(\rho)=e^{-\frac12 \omega\rho}(\omega \rho/2)^{\pm \frac{\sqrt{-A}}{2}}{}_1F_1\left(\frac12 \pm \frac{\sqrt{-A}}{2}-\frac{E}{4\omega},
1\pm \sqrt{-A}; \omega \rho\right),\ee
that are linearly independent as long as $1\pm \sqrt{-A}$ is not an integer \cite{gradshteyn2000table}. The requirement that the solution be continuous at the origin necessitates the choice of $R^+$ only. The requirement that $R(\rho)$ be square integrable 
implies that the ${}_1F_1$ must be a polynomial. This provides the quantization condition 
\be\fl \frac12 +\frac{\sqrt{-A}}{2}-\frac{E}{4\omega}=-m, \qquad i.e. \quad E=2\omega(2m+1+\sqrt{-A}), \qquad m \in \mathbb{Z}. \ee
The bound state eigenfunction $R(\rho)$ can alternatively be expressed in terms of the Laguerre polynomials as 
\be R(\rho)=e^{-\frac12 \omega \rho}(\omega \rho)^{\frac12 \sqrt{-A}}L_{m}^{\sqrt{-A}}(\omega \rho). \ee

Equation \eref{detS} can also be reduced to that for the confluent hypergeometric series by putting 
\be\fl  S(\sigma)=e^{\frac{\sqrt{-\alpha}\sigma^k}{2k}}\left(\frac{\sqrt{-\alpha}\sigma^k}{k}\right)^{\sqrt{-A}{2k}}{}_1F_1\left(\frac12-\frac{\beta}{4\sqrt{-\alpha}k}+\frac{\sqrt{-A}}{2k}, 1+\frac{\sqrt{-A}}{k}; \frac{\sqrt{-\alpha}\sigma^k}{k}\right),\label{SF} \ee
(the other solution is  singular at $\sigma =0$). Square integrability again necessitates that the ${}_1F_1$ functions should reduce to a polynomial and we obtain a quantization condition for the eigenvalue $A$, namely  $\frac 12 -\frac{\beta}{4\sqrt{-\alpha} k}+ \frac{\sqrt{-A}}{2k}=-n$ i.e.
\be A_n=-k^2\left(\frac{\beta}{2k\sqrt{-\alpha}}-2n-1\right)^2, \qquad \alpha<0, \beta>0.\ee
Note that $\alpha$ and $\beta$ satisfy the same bounds as in the classical case \eref{bounds}. 
We see that the quantized Hamiltonian \eref{HkM} only allows a finite number of bound states. Indeed, the condition $\sqrt{-A_n}>0$ implies 
\be n\leq \frac12 \left(\frac{\beta}{2k\sqrt{-\alpha}}-1\right)\label{nbounds}.\ee
In particular, the ground states exists only if 
\be \frac{\beta}{2k\sqrt{-\alpha}}\geq 1. \ee
In analogy with the classical case, the separation constant $A$ can take values only in a finite range (in this case a quantized one). Let us define $N$ as the largest integer satisfying \eref{nbounds} and define 
\be B=\frac{\beta}{2k\sqrt{-\alpha}}-2N-1.\ee
We then have 
\be \sqrt{-A_n}=k[B+2(N-n)].\ee
The functions $S(\sigma)$ of \eref{SF} can be expressed in terms of the Laguerre polynomials as 
\be\fl   S(\eta)=e^{-\frac{\eta}{2}}\eta^{\frac{B}{2}+N-n}L_n^{B+2(N-n)}(\eta), \qquad \eta=\frac{\sqrt{-\alpha}\sigma^k}{k}\label{Seta}.\ee
Changing variables in \eref{detS}, we see that $S(\eta)$ satisfies a self-adjoint differential equation, namely
\be \frac{\partial }{\partial \eta}\left(\eta \frac{\partial }{\partial \eta} S\right)+\left(\frac{1}{4k\eta}A+\frac14 \eta+\frac{\beta}{4k\sqrt{-\alpha}}\right)S=0.\ee
It follows that the eigenfunctions $S_n$ corresponding to different values of $n$ are mutually orthogonal 
\be \int_0^\infty S_nS_{n'}\frac{d\eta}{\eta}=N_n \delta_{n,n'}. \qquad 0\leq n,n'\leq N.\ee
We mention that this orthogonality does not follow from the orthogonality of the Laguerre polynomials $L_n^\alpha$ in \eref{Seta} since the index $\alpha$ depends on the lower index $n$. 

On the other hand, the functions $S(\sigma)$ solving \eref{detS} are directly related to the generalized Bessel polynomials $y_n(x,a,b)$ introduced by Krall and Fink \cite{krall1949new}. They satisfy the equation 
\be x^2y''+(ax+b)y'=n(n+a-1)y, \qquad y=y_n(x,a,b).\label{yn}\ee
Indeed, let us parameterize $\alpha$ in \eref{detS} as 
\be \alpha=-\frac{\beta^2}{4k^2(a-2)}, \qquad a\ne 2, \ee
change variables putting
 \be s=\frac{k}{\sqrt{-\alpha}\sigma^k}=\frac{1}{\eta}\ee
   and express the solution of \eref{detS} as 
   \be S_n(s)=s^{a-1}{2}e^{-\frac{1}{2s}}B_n(s,a, 1). \ee
   The polynomials $B_n(s, a,1)$ satisfy \eref{yn} and are hence the generalized Bessel polynomials defined as \cite{krall1949new, grosswald1978bessel} 
   \be B_n(s, a, b)=\sum_{k=0}^{n}C_{n,k}\ (n+k+a-2)^{(k)}\left(\frac{x}{b}\right)\ee
where $C_{n,k}$ is the binomial coefficient and 
\[ z^{(k)}=z(z-1)\ldots (z-k+1),\]
is a Pochhammer symbol. Note that the orthogonality of the Bessel polynomials over a finite range of indices is consistent with the results of \cite{fakhri2006ladder}.

\subsection{Superintegrability of the quantum system for rational k}
Let us now use the recurrence relation approach of Kalnins, Kress and Miller  \cite{KKM2011Recurr} to show that the quantum Hamiltonian is superintegrable. The main results of section \ref{52} are that the energy and separation constant for the system are
\bea E_{m,n}=2\omega\left(2m -2kn+1-k+\frac{\beta}{2\sqrt{-\alpha}}\right),\\
 \sqrt{-A_n}=k\left(-2n+\frac{\beta}{2\sqrt{-\alpha}k}-1\right) \eea
and the wave functions are given by 
\be \Psi_{m,n}=R_{m,n}(\xi)S_n(\eta),\ee
with
\bea R_{m,n}(\xi)=c_{m}e^{-\frac{\xi}{2}}\xi^{k(\frac{B}{2}+N-n)}L_m^{k(B+2N-2n)}(\xi), \qquad \xi=\omega \rho \label{Rmn}\\
 S_n(\eta)=d_{n}e^{-\eta/2}\eta^{\frac{B}{2}+N-n}L_n^{B+2N-2n}(\eta), \qquad \eta=\frac{\sqrt{-\alpha}\sigma^k}{k} \label{Sn}.\eea
  For rational $k=p/q$, we can exploit the degeneracy of the energy with respect to the transformations
\bea m\rightarrow m+p, \qquad n \rightarrow n+q\nn
m\rightarrow m-p, \qquad n \rightarrow n-q\nonumber\eea
to show that such a system is superintegrable. The additional integral of motion will be constructed from ladder operators for the associated Laguerre polynomials. 

As can be observed from \eref{Rmn} and \eref{Sn}, the appropriate ladder operators for the wavefunctions can be  constructed from ladder operators for the Laguerre polynomials that shift the degree by one and the parameter by two. To obtain such ladder operators, we take combinations of the following ladder and shift operators, see e.g. \cite{AAR}, 
\bea \frac{\partial}{\partial x}L_n^a(x)=-L_{n-1}^{a+1}(x)\label{Ld},\\
\left[x\frac{\partial}{\partial x}+(a-x)\right]L_n^a(x)=(n+1)L_{n+1}^{a-1}(x)\label{Lu},\\
\left[x\frac{\partial}{\partial x}+a\right]L_n^a(x)=(n+a)L_{n}^{a-1}(x)  \label{Sd},\\
\left[\frac{\partial}{\partial x}-1\right]L_n^a(x)=-L_n^{a+1}(x)\label{Su}.\eea
Taking \eref{Ld} composed with \eref{Su} and \eref{Lu} composed with \eref{Sd}, modulo the eigenvalue equation for the Laguerre polynomials 
\be \label{eigenLaguerre} \left[x\frac{\partial^2}{\partial x^2}+(1+a-x)\frac{\partial}{\partial x} +n\right]L_n^a(x)=0,\ee
yields first-order operators with the appropriate action on the Laguerre polynomials. Conjugating by an appropriate factor, gives the following ladder operators 
\bea \fl \left[\!\!(1+a)\frac{\partial}{\partial x} +\frac{2n+a+1}{2}-\frac{a(1+a)}{2x}\right]\!\!e^{-\frac{x}2}x^{\frac{a}{2}} L_n^a(x)\!=\!-e^{-\frac{x}2}x^{\frac{a}2+1}L_{n-1}^{a+2}(x)\label{Km}\\
 \fl \left[\!\!(1-a)\frac{\partial}{\partial x} +\frac{2n+a+1}{2}-\frac{a(a-1)}{2x}\right]\!\!e^{-\frac{x}2}x^{\frac{a}{2}}L_n^a(x)\!=\!(n\!+\! 1)(n\!+\!a)e^{-\frac{x}2}x^{\frac{a}2-1}L_{n+1}^{a-2}(x).\label{Kp}\eea

For the function $R_{m,n}(\xi)$, the ladder operators \eref{Km}, \eref{Kp} take the following forms
\bea K_{\sqrt{-A_n}, {E_{m,n}}}=\left(1+\sqrt{-A_n}\right)\partial_\xi-\frac{E_{m,n}}{4\omega} -\frac{1}{2\xi}\sqrt{-A_n}(1+\sqrt{-A_n})\\
K_{-\sqrt{-A_n},{E_{m,n}}}=\left(1-\sqrt{-A_n}\right)\partial_\xi-\frac{E_{m,n}}{4\omega} +\frac{1}{2\xi}\sqrt{-A_n}(1-\sqrt{-A_n}).\eea

The action of the K's on the functions $R_{m,n}$ is as follows 
\bea K_{\sqrt{-A_n},E_{m,n}}R_{m,n}=-R_{m-1, n-1/k}\nn
K_{-\sqrt{-A_n},E_{m,n}}R_{m,n}=-(m+1)(m+\sqrt{-A_n})R_{m+1, n+1/k}\nonumber.\eea
Thus, to raise or lower the index $m$ by $p$, we need only to apply the operator $p$ times
\bea K_{\sqrt{-A_n}, E_{m,n}}^p\equiv K_{\sqrt{-A_n}+2(p-1),E_{m,n}}\cdots K_{\sqrt{-A_n}+2,E_{m,n}}K_{\sqrt{-A_n}, E_{m,n}}\nn
 K_{-\sqrt{-A_n},E_{m,n}}^p\equiv K_{-\sqrt{-A_n}-2(p-1),E_{m,n}}\cdots K_{-\sqrt{-A_n}-2,E_{m,n}}K_{-\sqrt{-A_n},E_{m,n}}\nonumber.\eea
Note that after each successive application, we change the operator by adding 2 to $\sqrt{-A_n}$ but the quantity $E_{m,n}$ is unchanged in each iteration. 
Note that by adding or subtracting $p$ from $\sqrt{-A_n}$ we change $n$ by $q$
\be \sqrt{-A_n}\pm 2p=\frac{p}{q}\left(-2n+\frac{\beta}{2\sqrt{-\alpha}k}-1\right)\pm 2p=\sqrt{-A_{n\mp q}},\ee
and so the action of these operators $K^p$ is given by 
\bea K_{\sqrt{-A_n},E_{m,n}}^pR_{m,n}=(-1)^pR_{m+p,n+q}\\
     K_{-\sqrt{-A_n},E_{m,n}}^p R_{m,n}=(m+1)_p(1-m-\sqrt{-A_n})_p R_{m-p, n-q}.\eea

Similarly, we have raising and lowering operators for the functions $S_n(\eta)$ given by 
\bea \fl  J_{\sqrt{-A_n}/k}=\left(1+\sqrt{-A_n}/k\right)\partial_\eta-\frac{\beta}{4\sqrt{-\alpha}k} -\frac{1}{2k^2\eta}\sqrt{-A_n}(k+\sqrt{-A_n})\\\
\fl J_{-\sqrt{-A_n}/k }=\left(1-\sqrt{-A_n}/k\right)\partial_\eta-\frac{\beta}{4\sqrt{-\alpha}k} +\frac{1}{2k^2\eta}\sqrt{-A_n}(k-\sqrt{-A_n}).\eea
The action on the basis is 
\bea J_{ \sqrt{-A_n}/k}S_n=-S_{n-1}\nn
     J_{-\sqrt{-A_n}/k}S_n=-(n+1)(-n+\frac{\beta}{2\sqrt{-\alpha}k}-1)S_{n+1}\nonumber.\eea
Again, the repeated application of the ladder operators is defined as 
\bea J_{ \sqrt{-A_n}/k}^q\equiv J_{ \sqrt{-A_n}/k+2(q-1)}\cdots J_{ \sqrt{-A_n}/k+2}J_{ \sqrt{-A_n}/k,b}\nn
     J_{-\sqrt{-A_n}/k,b}^q\equiv J_{-\sqrt{-A_n}/k-2(q-1)}\cdots J_{-\sqrt{-A_n}/k-2}J_{-\sqrt{-A_n}/k},\nonumber\eea 
with action on  $S_n$ as 
\bea J_{ \sqrt{-A_n}/k}^qS_n=(-1)^qS_{n-q}\\
    J_{-\sqrt{-A_n}/k}^qS_n=(n+1)_q(n-\frac{\beta}{2\sqrt{-\alpha}k})_q  S_{n+q}.\eea

Thus, we have created parameter dependent raising and lowering operators which map between wave fucntions with the same energy values given by 
\bea T_1=K_{\sqrt{-A_n},a}^pJ_{\sqrt{-A_n}/k,b}^q\\
T_2= K_{-\sqrt{-A_n},a}^pJ_{-\sqrt{-A_n}/k,b}^q.\eea
To remove the dependence of these operators on the indices $m,n$, first we push the constant $E_{m,n}$ to the right and then replace with the operator $H$. 
Thus, when the resulting operator acts on a wave function $\Psi_{m,n}$, it takes the value of the energy $E_{m,n}$
Next, we note that after this replacement, the resulting operators
\be T_s=T_1+T_2, \qquad T_{a}=\frac{1}{\sqrt{-A_n}}(T_1-T_2),\ee
are polynomials in $A_n$ (and hence the index $n$) and so we can again push this constant to the right and replace it with the first integral of the motion
\[ A_n\rightarrow  L_1=- \partial_s^2-\alpha e^{4ks} -\beta e^{2ks}\]
to remove the dependence of the operator on $n$. 
In this way, we have removed the dependence of the operators on the quantum numbers $m$ and $n$ and so 
\be [T_s, H_k] \Psi_{m,n}=[T_a, H_k]\Psi_{m,n}=0, \qquad \forall m,n.\ee
By a standard  arguments as given in \cite{KKM2011Recurr}, the identities hold in general (not only on the solutions of the Schr\"odinger equation) 
\be  [T_s, H_k] =[T_a, H_k]=0,\ee
Thus, we have proven that the system is superintegrable. 

\section{Conclusions}
 In Section 2 we have proposed a method for expanding an integrable and exactly solvable system that separates in subgroup type coordinates into an infinite family of such systems. In the article, we have shown that in two-dimensional flat spaces the expanded family remains in the same space ($E_2(\mathbb{C}), \, E_2$ or $E_{1,1}$) and that the method also preserves superintegrability. The method can also be used in higher-dimensional spaces and in non-flat ones. It will always embed a given integrable and exactly solvable system into a parameterized family of such systems. In general, it will however not preserve the structure of the space, i.e. it can change the metric. There is also no guarantee that when applied to a given superintegrable system, the method will preserve superintegrability. 
 
 In Sections 3 and 4, the superintegrability of the classical and quantum system is demonstrated. For the classical system, the additional integral of the motion was constructed from the bounded trajectories thus proving the converse of the theorem of Nekhoroshev \cite{nekhoroshev1972action} for this system. For the quantum system, the exact-solvability of the wave-functions was used to construct the additional integral, thus proving the converse of the conjecture of Tempesta, Turbiner and Winternitz \cite{TempTW} for this system. 
 
 The exact-solvability of the quantum system \eref{HkM} in terms of generalized Bessel polynomials, provides yet another instance of the deep connection between superintegrable systems and orthogonal polynomials. Beyond the classical families of orthogonal polynomials known to be associated with superintegrable systems, there have been recent advances in discovering the connection between superintegrable systems and new families of orthogonal polynomials. In particular, new families of superintegrable systems associated with the $q\rightarrow -1$ limit of q-Jacobi polynomials \cite{PVZ2011inffam} and with exceptional Jacobi polynomials \cite{post2012families}. The connection between these two fields is evidenced in the methods of constructing the integrals of the motion from the ladder operators for the polynomials, as above and in \cite{KKM2011Recurr, marquette2010superintegrability, marquette2011infinite}, as well as in the representation theory of the algebras generated by  such integrals see e.g.  \cite{kalnins2011two, Post2011}. Expanding the connection even further, are recent results linking other special functions, including elliptic and Painlev\'e functions, to superintegrable systems, see e.g. \cite{Gravel, GW, MW2008, marquette2009painleve, marquette2011infinite, TW20101}.

\appendix

\section{Trajectories of the  harmonic oscillator in $E_{1,1}$}\label{SHO}
We give the trajectories of the harmonic oscillator in $E_{1,1}$, which were previously studied in \cite{pavsic1999pseudo}. In Cartesian coordinates the Hamiltonian is simply the difference of two one-dimensional oscillators 
\be H^{O}=p_u^2+\omega^2 u^2-p_v^2-\omega^2v^2.\ee
The trajectories, up to rotation of the plane and translations in time, are given by 
\bea u(t)=a\sin 2\omega t, \qquad &v(t)=b \cos 2 \omega t, \\
     p_u(t)=\omega a \cos 2\omega t, \qquad &p_v(t)=-b\omega \sin 2 \omega t.\eea
The energy can take positive and negative values and is defined by
\be E=\omega^2(a^2-b^2).\ee
In the coordinate system $(\rho, \sigma)$ the trajectories are given by 
\bea\label{rhosho} \rho(t)=a^2\sin^2 2\omega t-b^2 \cos 2 \omega t,\nn
\sigma(t)=\frac{a \sin 2 \omega t+ b \cos 2 \omega t}{a \sin 2 \omega t-b \cos 2 \omega t},\nn
p_\rho(t)=\frac{\omega (a^2+b^2) \sin 2\omega t \cos 2\omega t}{2(a^2\sin^2 2\omega t-b^2 \cos 2 \omega t)}\nn
p_\sigma(t)=\frac{\omega a b(a\sin 2\omega t-b\cos 2 \omega t)}{2(a\sin 2\omega t+b\cos 2 \omega t)}
\eea
with the  separation constant $ A\equiv 4 \sigma^2 p_\sigma^2$ taking values $A=\omega^2 a^2b^2.$ 

\ack { The authors would like to thank Willard Miller Jr. for helpful discussions and for bringing relevant papers to our attention. S. P. acknowledges a postdoctoral fellowship awarded by the Laboratory of Mathematical Physics of the Centre de Recherches Math\'ematiques, Universit\'e de Montr\'eal. The research of P. W. is partially supported by a research grant from NSERC of Canada.}

 \section*{References}
 \bibliography{all}
 \bibliographystyle{jphysa}

\end{document}